\documentclass[useAMS,usenatbib,twocolumn]{mnras}
\usepackage{graphicx,url,amssymb,amsmath,color,units,wasysym,epsfig,epstopdf,enumerate,tabularx, subfigure,hyperref}
 
\usepackage{newtxtext,newtxmath}
\usepackage[T1]{fontenc}
\usepackage{ae,aecompl}

\usepackage[dvipsnames]{xcolor}
\usepackage[normalem]{ulem}
\usepackage[bottom]{footmisc}

\renewcommand{\sc}[1]{\textsc{#1}}

\colorlet{darkorange}{black!30!orange!70!}

\begin{document}

\title[Argument of periapsis]{Gravitational-wave inference for eccentric binaries: the argument of periapsis}

\author[T. A. Clarke et al.]{
Teagan A. Clarke$^{1,2}$\thanks{teagan.clarke@monash.edu}, Isobel M. Romero-Shaw$^{1,2,3}$, Paul D. Lasky$^{1,2}$ and Eric Thrane$^{1,2}$\\
$^{1}$School of Physics and Astronomy, Monash University, VIC 3800, Australia\\
$^{2}$OzGrav: The ARC Centre of Excellence for Gravitational-wave Discovery, Clayton, VIC 3800, Australia\\
$^{3}$Department of Applied Mathematics and Theoretical Physics, Cambridge CB3 0WA, United Kingdom
}

\date{Accepted 2022 October 11. Received 2022 September 09; in original form 2022 June 29}

\pubyear{2022}

\maketitle

\begin{abstract}
Gravitational waves from binary black hole mergers have allowed us to directly observe stellar-mass black hole binaries for the first time, and therefore explore their formation channels. One of the ways to infer how a binary system is assembled is by measuring the system's orbital eccentricity. Current methods of parameter estimation do not include all physical effects of eccentric systems such as spin-induced precession, higher-order modes, and the initial argument of periapsis: an angle describing the orientation of the orbital ellipse. We explore how varying the argument of periapsis changes gravitational waveforms and study its effect on the inference of astrophysical parameters. We use the eccentric spin-aligned waveforms \texttt{TEOBResumS} and \texttt{SEOBNRE} to measure the change in the waveforms as the argument of periapsis is changed. We find that the argument of periapsis could already be impacting analyses performed with \texttt{TEOBResumS}. However, it is likely to be well-resolvable in the foreseeable future only for the loudest events observed by LIGO--Virgo--KAGRA. The systematic error in previous, low-eccentricity analyses that have not considered the argument of periapsis is likely to be small. 
\end{abstract}


\section{Introduction}
Approximately 90 gravitational-wave events have been detected \citep{gwtc3,O3b_population}, including two binary neutron star mergers \citep{GW170817, 2020_GW190425}, approximately two neutron star-black hole binary mergers \citep{nsbh_detection} and over 80 binary black hole mergers \citep{GWTC-1, GWTC-2, gwtc3}. Despite the wealth of observations, the question of how the population of merging compact binaries formed has proved challenging to answer. For stellar-mass binary black holes, there are two overarching formation channels that could result in coalescence within the Hubble time: isolated binary evolution and dynamical assembly (for a recent overview see, e.g., \cite{Mandel_2022}). A binary black hole formed in isolation undergoes normal binary stellar evolution until both stars collapse into black holes, with no interaction with external objects \citep[e.g.,][]{Bethe98, belczynski_2016, Stevenson_2017_nat}. Alternatively, black hole binaries may form via dynamical assembly: both objects have already evolved into black holes, and become gravitationally bound in a densely-populated environment like a globular cluster or galactic nucleus \citep[e.g.,][]{OLeary05, rodriguez_2016_gc, Yang2019, Grobner20}. 

Measuring the orbital eccentricity, along with the component masses and spins, of the black holes in a binary can help determine how the binary formed. Since binaries circularise through the emission of gravitational waves \citep{Peters64}, isolated binaries are expected to have almost circular orbits when they enter the observing band of LIGO--Virgo--KAGRA (LVK) $\approx\unit[10-2000]{Hz}$ \citep{adv_ligo_2015, AdvancedVirgo, 2020_Kagra}. Meanwhile, binaries formed through dynamical assembly can merge very quickly, and hence maintain measurable eccentricity in the LVK observing band \citep[e.g.,][]{Oleary2009, Rodriguez18b, Zevin:2021:seleccentricity}. 

Signatures of dynamical formation, such as orbital eccentricity \citep{Romero-Shaw:2019itr, Lower18, Romero-Shaw2021, romero_shaw_22, OSheaKumar2021} and misaligned spins \citep{GWTC-2_RnP}, inferred in some of the existing gravitational-wave observations suggest that dynamically-formed systems may make up a substantial sub-population of binary black holes that merge.
However, at least some binaries must be assembled in the field to account for the tendency of LVK binaries to merge with aligned spin \citep{O3b_population, Tong_2022}. Up to four of the binary black hole mergers observed to date have been identified as potentially eccentric \citep{Romero-Shaw:2019itr, Wu_2020, Romero-Shaw2021, OSheaKumar2021, romero_shaw_22, Iglesias_2022}, including the high-mass system, GW190521 \citep{GW190521-detection, Romero-Shaw:2020:GW190521,Gamba_2021, Gayathri_2022}.

\begin{figure*}
    \centering
    \includegraphics[trim = 30mm 0mm 30mm 0mm, scale=0.55]{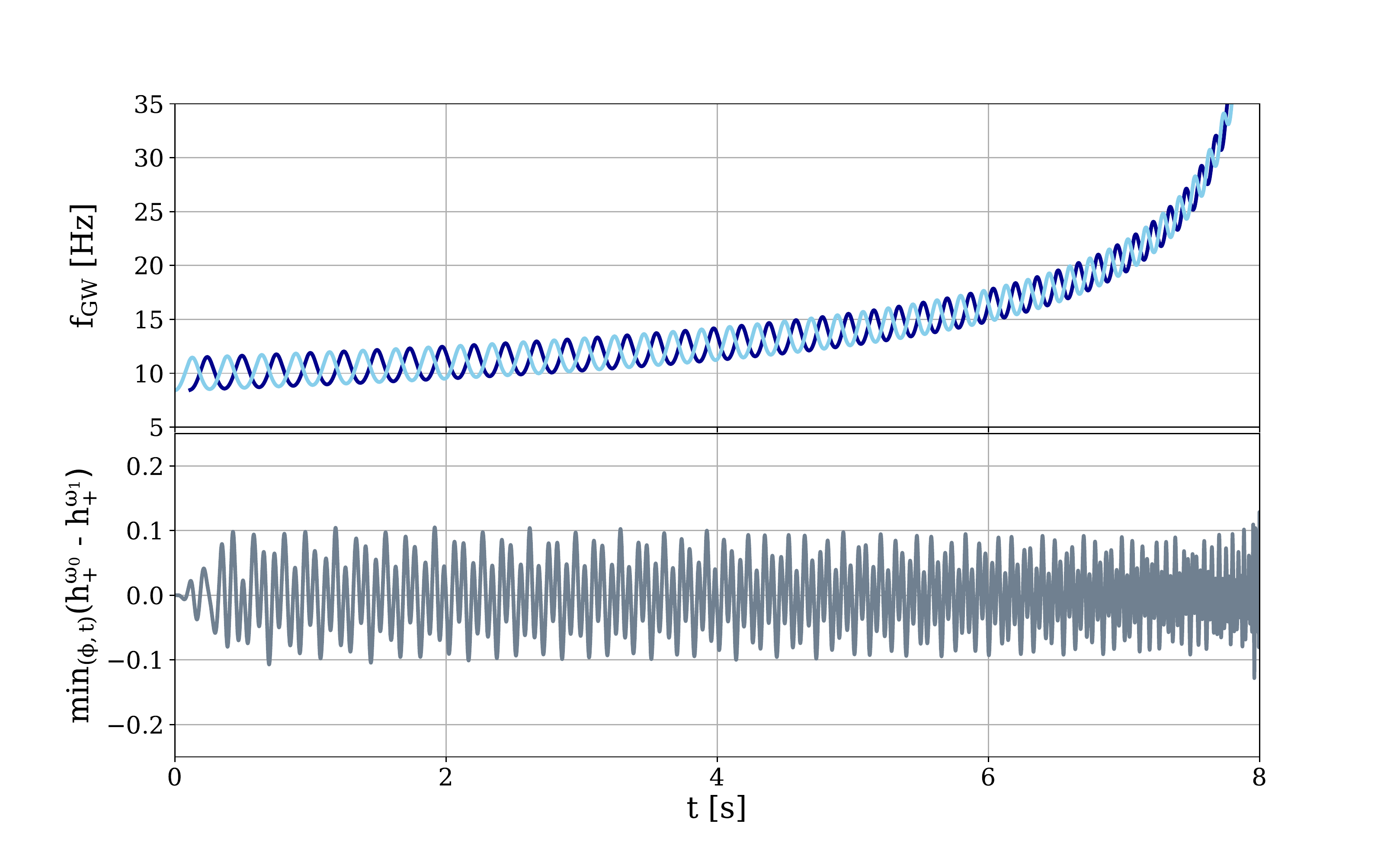}
    \caption{The effect of changing the reference argument of periapsis $\omega_\text{ref}$ on the gravitational waveforms for a GW150914-like event with parameters listed in Table~\ref{tab:teob_params}. Top panel: the frequency evolution of a binary black hole inspiral of f$_\text{ref}$ = $10$~Hz, e$_\text{ref}$ = 0.1 and f$_\text{ref} = 9.95$~Hz, e$_\text{ref}$ = 0.1005. This corresponds to a change in $\omega_\text{ref}$ of $\approx \pi$. It can be thought of as the system being evolved back in time by half a waveform cycle and illustrates the way waveforms change when they have different values of $\omega_\text{ref}$. The peaks and troughs show the periapsis and apoapsis passages respectively. The change in frequency in one waveform cycle is approximated as the difference in frequency between two troughs. Bottom Panel: The difference between the waveforms plotted in the top panel, minimised over phase and time.}
    \label{fig:frequency}
\end{figure*}

While most of the discussion of eccentric binaries has focused on measuring the eccentricity $e$, the gravitational-wave signal from an eccentric binary is also affected by the argument of periapsis at the reference frequency $\omega_\text{ref}$. This parameter is the angle of rotation of an elliptical orbit relative to a reference plane, and is one of the 17 parameters that fully describe an eccentric binary black hole system.
To illustrate how the argument of periapsis affects gravitational waveforms, we plot in Fig. \ref{fig:frequency}, the gravitational-wave frequency of the dominant ($\ell=2,|m|=2$) mode as a function of time for two gravitational waveforms, generated with a difference of $\pi$ in their reference $\omega$.
This change causes the binary to experience periapsis and apoapsis at different frequencies, despite them having the same eccentricity at $\unit[10]{Hz}$.

There are few eccentric waveform approximants available and none currently used for astrophysical inference of LVK data include $\omega_\text{ref}$ as a parameter. The waveform models used to search for eccentricity in the studies above---\texttt{SEOBNRE} \citep{SEOBNRE, SEOBNREfrequencydomain, validationSEOBNRE}, \texttt{TEOBResumS} \citep{2018_Teob, Teob_2020, TEOBResumS} and \texttt{EccentricFD} \citep{eccentricFD}---do not allow allow the user to straightforwardly vary $\omega_\text{ref}$. However, some new eccentric waveform models are intended to provide this option. \cite{Klein2021} has developed a new waveform model that allows the user to vary $\omega$; this waveform model has been used to study inference of eccentric binaries with LISA \citep{buscicchio_2021}.

\cite{Islam:2021:meananomaly} developed a numerical relativity surrogate waveform with a variable mean anomaly $l_\text{ref}$. By simulating equal mass binary, moderately eccentric ($e_\text{ref} = 0.1$) waveform predictions with varied $l_\text{ref}$ in white noise, they found waveform mismatches up to 0.1. This result would seem to suggest that the argument of periapsis has an important effect on the eccentricity measurements obtained using current eccentric waveforms.  However, the effect of $\omega_\text{ref}$ on gravitational-wave inference is poorly understood since previous analyses used a fixed value of $\omega_\text{ref}$, set by the choice of starting eccentricity and reference frequency. Understanding the role of $\omega_\text{ref}$ is important to avoid bias in eccentric parameter estimation. Systematic error related to the argument of periapsis has been assumed to be small, but in light of work by \cite{Islam:2021:meananomaly}, this assumption must be checked \citep{Lower18, Romero-Shaw2021}. In this paper, we investigate the effect of $\omega_\text{ref}$ on gravitational-wave source inference. In Section \ref{sec:omega} we describe the argument of periapsis and our prescription for measuring it in Section \ref{sec:method}. In Section \ref{sec:results}, we assess the extent to which $\omega_\text{ref}$ can be resolved and discuss the implications of our results.

\section{The effective Argument of Periapsis}
\label{sec:omega}
To investigate the effect of the reference argument of periapsis $\omega_\text{ref}$ on gravitational-wave source inference, we must find a way to change $\omega_\text{ref}$ in the waveform models so that we may measure the effect it has on parameter estimation. Unfortunately, no currently available waveform approximants allow the user to directly control $\omega_\text{ref}$. Thus, in this section, we devise a mechanism that we can use to vary $\omega_\text{ref}$ indirectly. We use two waveform models for our demonstration: the time-domain effective one-body \citep{Buonanno_1999} waveform models \texttt{TEOBResumS} \citep{2018_Teob, Teob_2020, TEOBResumS} and \texttt{SEOBNRE} \citep{SEOBNRE, SEOBNREfrequencydomain, validationSEOBNRE}. The eccentric version of \texttt{TEOBResumS} is validated against numerical relativity simulations for eccentricities up to $e \lesssim 0.3$ at $10$~Hz for a binary with total mass $60$~M$_\odot$. \texttt{SEOBNRE} is validated up to eccentricities of 0.2 at $10$~Hz. Both waveforms are limited to spin-aligned systems.\footnote{Comparing the eccentricities of the two waveforms is not straightforward, since they employ different definitions of eccentricity and its evolution with frequency. We ignore this added complication for this study, however future work should consider how eccentricity values map to each other in different waveforms. This is explored by \cite{knee22}.}

Using these waveforms, $\omega_\text{ref}$ can be indirectly varied by changing the waveform reference frequency $f_{\text{ref}}$, and the eccentricity at $f_\text{ref}$, $e_{\text{ref}}$. By setting these waveform parameters, we set an unknown but specific $\omega_\text{ref}$, which is the argument of periapsis at $f_\text{ref}$. The variable $\omega_\text{ref}$ should change by $2\pi$ when the reference frequency and eccentricity have been varied through one orbital period. This means that we can vary $\omega_\text{ref}$ indirectly by following the waveform through a cycle of eccentricity and frequency evolution.
The trick is to find the path through $(e_\text{ref}, f_\text{ref})$ that corresponds to a fixed value of eccentricity at $\unit[10]{Hz}$.
We call this ``the $e_{10}$ path''.
Each point along the $e_{10}$ path corresponds to a different value of $\omega_\text{ref}$.

In order to estimate the $e_{10}$ path, we evolve the orbital eccentricity from $f_\text{ref}$ back to $\unit[10]{Hz}$.
To this end we employ a post-Newtonian approximation that describes the eccentricity as a function of gravitational-wave frequency. We use the approximation outlined in \cite{Moore_2016}, who show that the eccentricity as a function of frequency can be calculated analytically to 3PN order if the eccentricity is assumed to be small (better than 2\% at $e=0.1$ for low frequencies $\lesssim \unit[200]{Hz}$): 
\begin{equation}
    e_t = e_0 \Big(\frac{\xi_{\phi_0}}{\xi_{\phi}}\Big)^{\frac{19}{18}} \ \frac{\epsilon(\xi_\phi)}{\epsilon(\xi_{\phi_0})} ,
\label{eq:e_t}
\end{equation}
where 
\begin{equation}
\xi_\phi = (m_1+m_2)\pi f_\text{GW},
\end{equation}
is a dimensionless frequency parameter, which serves as the PN expansion parameter, and $\epsilon (\xi_\phi) $ is a 3PN correction term. From this \eqref{eq:e_t} becomes, at 0th order, in the $e \rightarrow 0$ limit:
\begin{equation}
    e_t(f_\mathrm{GW}) \approx e_0 \Big( \frac{f_\mathrm{GW_t}}{f_\mathrm{GW_0}}\Big)^{-19/18}. 
\label{eq:e_t2}
\end{equation}
We use this equation to trace out the $e_{10}$ path from $f_\text{ref}$ to $\unit[10]{Hz}$. As we move along the $e_{10}$ path, we vary $\omega_\text{ref}$ ---our indirect estimate of the reference argument of periapsis. By studying how the waveform changes for different values of $\omega_\text{ref}$ along the $e_{10}$ path, we can assess the affect of the argument of periapsis on gravitational-wave inference. 

The waveform overlap \citep{flanagan1998} describes the similarity between two gravitational waveforms. By calculating the overlap between waveforms that are the same in all parameters besides the argument of periapsis, we can quantify the amount $\omega_\text{ref}$ changes the waveforms. The phase and time maximised overlap is given by
\begin{equation}
    \mathcal{O} = \mathrm{max}(\phi_0, t_0) \frac{\langle h_1 | h_2 \rangle}{\sqrt{\langle h_1|h_1 \rangle \langle h_2 | h_2 \rangle}}, 
    \label{eq:overlap}
\end{equation}
where $\langle a | b \rangle$ is the inner product defined such that
  
  \begin{equation}
        \langle a | b \rangle = 4 \mathrm{Re} \int ^\infty _0 df \frac{\tilde{a}(f)\tilde{b}(f)}{S_h(f)} ,
        \label{eq:inner_product}
    \end{equation}
    where $S_h(f)$ is the power spectral density of the noise. We calculate the overlap over a grid of waveforms generated with \texttt{TEOBResumS} and \texttt{SEOBNRE}, corresponding to the predicted change in eccentricity and frequency over an orbital cycle.
    
\begin{figure}
    \centering
    \includegraphics[trim=10mm 0mm 10mm 0mm, width = \columnwidth]{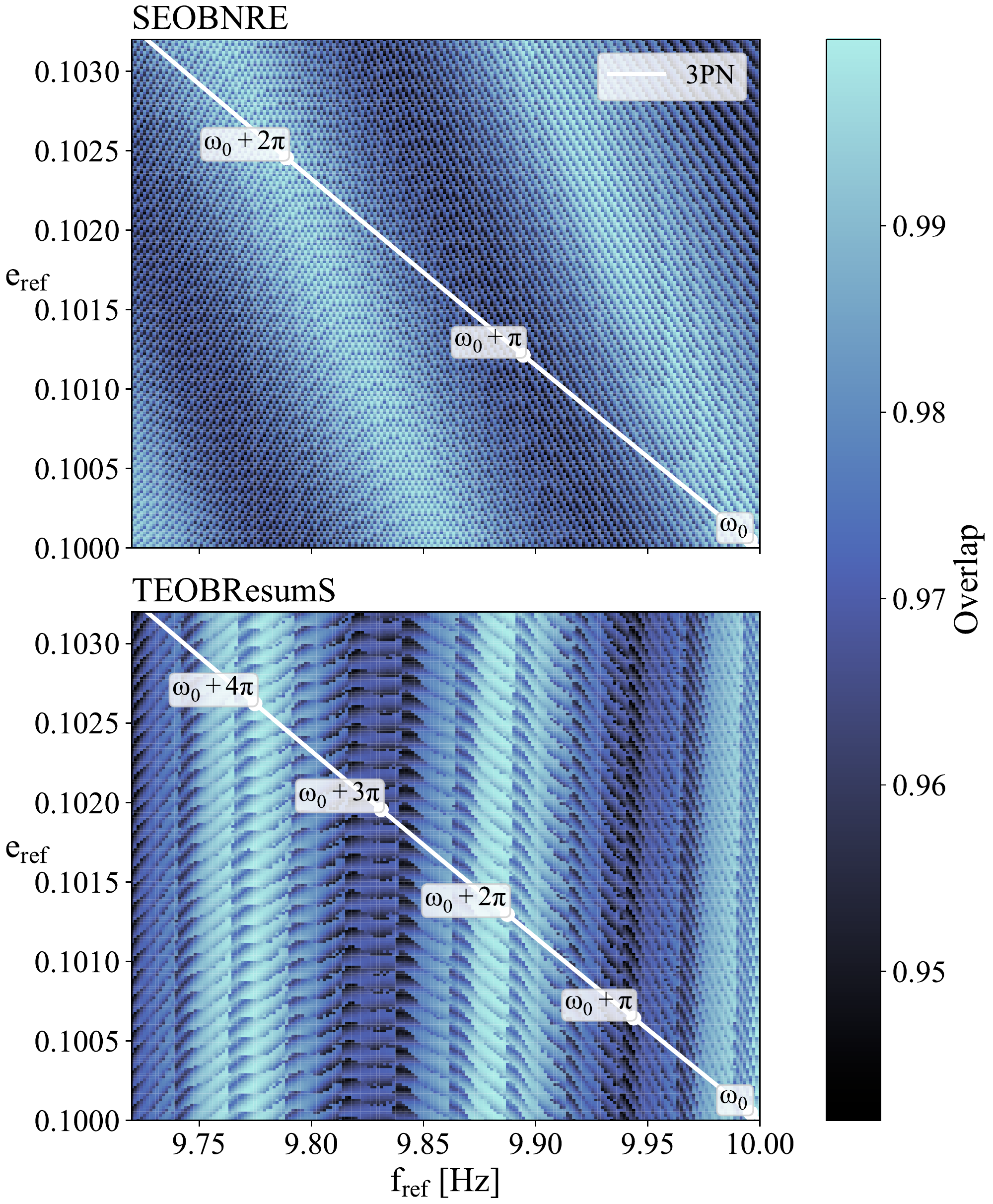}
    \caption{The maximised overlap over approximately two waveform periods. Top panel: \texttt{SEOBNRE}. Bottom panel: \texttt{TEOBResumS}. The white line shown is the 3PN expansion, showing the evolution of a system starting at the fiducial waveform, labelled as $\omega_0$ in the bottom right corner. The overlap varies by about 0.05 throughout the waveform cycle, suggesting that $\omega_\text{ref}$ has a non-negligible effect on the waveforms. The `stripes' of  constant overlap exhibit different orientations for \texttt{SEOBNRE} and \texttt{TEOBResumS}. We speculate that this is due to differing waveform systematics, particularly different definitions of eccentricity employed in the waveforms \citep[see e.g., ][]{knee22}.}
    \label{fig:overlap_1_cycle}
\end{figure}

    We generate the reference waveform with the parameters listed in Table \ref{tab:teob_params}. The overlap $\mathcal{O}=1$ when the reference waveform and comparison waveform are the same. We calculate the overlap (maximized over phase and time) on a grid of $(e_\text{ref}, f_\text{ref})$. Each grid-space records the overlap between a waveform with $(e_\text{ref}, f_\text{ref})$ and the fiducial waveform at $e_{\text{fiducial}}=0.1, f_{\text{fiducial}} = \unit[10]{Hz}$. This is shown in Fig. \ref{fig:overlap_1_cycle} along with the $e_{10}$ path. The effective argument of periapsis $\omega_\text{eff}$ parameterises the location along the $e_{10}$ path and our measurement of $\omega_\text{ref}$. We assume that $\omega_\text{eff}$ values are evenly spaced along the curve and that $\omega_\text{eff} = 0$ when $e_\text{ref}, f_\text{ref} = e_\text{fiducial}, f_\text{fiducial}$. In reality $\omega_\text{ref}$ at the fiducial waveform is arbitrary. The waveform overlap follows a sinusoidal pattern and reaches a minimum of $\approx 0.95$ at $\omega_\text{eff} \approx \pi$. This is expected because when $\omega_\text{eff} = 2\pi$, the waveforms are the same but initialised one cycle apart, resulting in a local maximum for the waveform overlap. Our change in overlap does not match that of \cite{Islam:2021:meananomaly}, potentially because we used LVK noise, rather than white noise. This suggests that $\omega_\text{ref}$ is less resolvable in realistic (noisy) gravitational-wave data.
    
\begin{table}
\caption{Source parameters of the injected fiducial waveform used to calculate the grid of overlaps and likelihood of $\omega_\text{ref}$ using \texttt{TEOBResumS}.  The black hole masses are measured in the detector frame. The parameters are chosen to be similar to GW150914, but the distance increased such that the SNR $\approx$ 17 at design sensitivity.}
\begin{center}
\begin{tabular}{c  c  c } 
 \hline
 Parameter & Abbreviation & Value  \\ [0.5ex] 
 \hline
 black hole masses & m$_1$, m$_2$ & 30, 25 M$_\odot$  \\ 
 
 reference eccentricity & e$_{\text{fiducial}}$ & 0.1  \\
 
 reference frequency & f${_\text{fiducial}}$ & $10$~Hz \\
 
 spin parameters & $\chi_1$,$\chi_2$ & 0.0 \\
 
 inclination & $\theta_\text{JN}$ & 0.6  \\
 
 phase & $\phi_c$ & 1.5 \\
 
 luminosity distance & D$_L$ & 1419 Mpc \\
 
 right ascension & RA & 3.5 \\

 declination & Dec & 0.5 \\
 \hline
\end{tabular}
\label{tab:teob_params}
\end{center}
\end{table}



 
 
 
 



Two waveforms with different reference arguments of periapsis (but otherwise identical parameters) can be distinguished from the reference waveform if \citep{lindbolm_2008, baird_2013}:
\begin{equation}
   1- \mathcal{O} \sim \mathrm{SNR^{-2}},
   \label{eq:SNR}
\end{equation}
where SNR $= \sqrt{\langle h_0|h_0\rangle}$ is the optimal matched-filter SNR of the reference waveform $h_0$. Hence, for the lowest value of the overlap in Fig. \ref{fig:overlap_1_cycle} of $\approx 0.95$, an SNR of $\approx$ 5 is required to distinguish this waveform from the reference waveform. Of course, this assumes that all other parameters are known perfectly, which is not the case for real inference calculations in noisy data. 
In the subsequent section, we determine the extent to which $\omega_\text{ref}$ can be resolved in noisy data.

\section{Method}
\label{sec:method}
We calculate the posterior distribution for $\omega_\text{ref}$ for a simulated eccentric gravitational-wave signal. 
Table~\ref{tab:teob_params} shows the injection parameters of the chosen eccentric fiducial waveform. We choose the system to have a relatively loud but realistic SNR of $\approx$ 17 and parameters similar to GW150914. These parameters are chosen to allow for easier comparison with other studies that focus on GW150914-like events. While future studies should investigate the effect of $\omega$ on high-mass systems such as GW190521, we only study GW150914-like systems because it is difficult to use our method to approximate the e$_{10}$ path for massive systems that spend very little time in-band. The first step is to generate standard posterior samples at a fixed value of $f_\text{ref}$.
To this end, we carry out parameter estimation using the Bayesian Inference Library (\texttt{Bilby}) and the \texttt{bilby$\_$pipe} pipeline \citep{bilby, Romero-Shaw:2020:Bilby}, the spin-aligned eccentric waveform approximant \texttt{TEOBResumS} \citep{TEOBResumS} and the nested sampler \texttt{dynesty} \citep{dynesty}.\footnote{We implement the speed-up trick described in \cite{OSheaKumar2021}, where the integrator error tolerances are loosened slightly. This modification allows full parameter estimation to be performed directly with \texttt{TEOBResumS}.} We perform an additional sampling run at SNR $\approx$ 30 to compare the results. We also generate posterior samples injected and recovered with \texttt{SEOBNRE} \citep{SEOBNRE} for an injection with similar parameters to \texttt{TEOBResumS} at SNR $\approx$ 17.\footnote{\texttt{SEOBNRE} samples are generated by performing likelihood reweighting \citep{Payne2019} on samples generated with the fast quasi-circular waveform \texttt{IMRPhenomD} \citep{Khan15}. This has the disadvantage of reducing the number of effective posterior samples.} We inject signals into Gaussian noise coloured by the LIGO amplitude spectral density noise curves at design sensitivity.\footnote{amplitude spectral density curves are taken from https://dcc.ligo.org/LIGO-T2000012/public \citep{ALIGO}} We use uniform priors in the component masses, spins, luminosity distance and eccentricity.\footnote{We sample with 1000 live points, phase and time marginalisation turned on and a stopping criterion of $ \Delta \text{log}\mathcal{Z} < 0.1$, where $\mathcal{Z}$ is the Bayesian evidence.}
The initial posterior samples from this step are all (inadvertently) assigned some \textit{implicit} argument of periapsis $\omega_\text{fixed}$, which is completely determined by $(e_\text{ref}, f_\text{ref})$.
In this sense, $\omega_\text{ref}$ is not a free parameter of the initial posterior samples.

The next step is to importance sample the initial posterior samples in order to obtain the results we would have obtained if $\omega_\text{ref}$ had been a free parameter.
For each sample $i$, we calculate a weight
\begin{align}\label{eq:weight}
    w(\theta_i | d) = \frac{\int d\omega \,  {\cal L}(d|\theta_i, \omega) \, \pi(\omega) }{{\cal L}(d|\theta_i, \omega_\text{fixed})} .
\end{align}
The numerator of the weight is the ``target likelihood'' that marginalises over $\omega_\text{ref}$ while the denominator is the ``proposal likelihood'' used to generate the initial samples.\footnote{When calculating the weight in Eq.~\ref{eq:weight}, both likelihoods are implicitly marginalised over the time and phase of coalescence.}
The numerator integral over $\omega_\text{ref}$ is along the $e_{10}$ path described above.
Next, for each sample $i$, we calculate the posterior probability density for $\omega_\text{ref}$ given parameters $\theta_i$: $p(\omega | d, \theta_i)$; see the light blue traces in Fig.~\ref{fig:likelihood}, which are proportional to $\ln p(\omega | d, \theta_i)$.
We use the weights $w(\theta_i|d)$ and the posteriors $p(\omega | d, \theta_i)$ to calculate the posterior probability density of $\omega_\text{ref}$ given the data:
\begin{align}
    p(\omega | d) \propto & \ \pi(\omega) \, {\cal L}(d | \omega) \nonumber\\
    \propto & \int d\theta \, \mathcal{L}(d | \theta, \omega) \, \pi(\theta) \nonumber\\
    \propto & \int d\theta \, \pi(\theta)
   \left(\frac{\mathcal{L}(d | \theta, \omega)}{\mathcal{L}(d | \theta, \omega_\text{fixed})}\right)
    \mathcal{L}(d | \theta, \omega_\text{fixed}) \nonumber\\
     \propto & \int d\theta \, \pi(\theta)
    \left(\frac{p(\omega | d, \theta) \int d\omega' \, \mathcal{L}(d | \theta, \omega') }{\mathcal{L}(d | \theta, \omega_\text{fixed})}\right)
    \mathcal{L}(d | \theta, \omega_\text{fixed}) \nonumber\\
    \propto & \int d\theta \, 
    p(\omega | d, \theta) \, 
    w(\theta | d) \, 
    \bigg(
    \pi(\theta) \, 
    \mathcal{L}(d | \theta, \omega_\text{fixed}) \bigg) \nonumber\\
    \propto & \sum_i \, 
    p(\omega | d, \theta_i) \, 
    w(\theta_i | d) \nonumber\\
\label{eq:posterior}
\end{align}
In this derivation we implicitly assume a uniform prior for $\omega_\text{ref}$.
In the final line, the posterior is written as a sum over initial samples.
Graphically, this implies that the posterior for $\omega_\text{ref}$ is a weighted average of the (exponential of the) light blue curves in Fig.~\ref{fig:likelihood}.

\section{Results and Discussion}
\label{sec:results}
\begin{figure}
    \centering
    \includegraphics[trim = 15mm 15mm 15mm 15mm,width=\columnwidth]{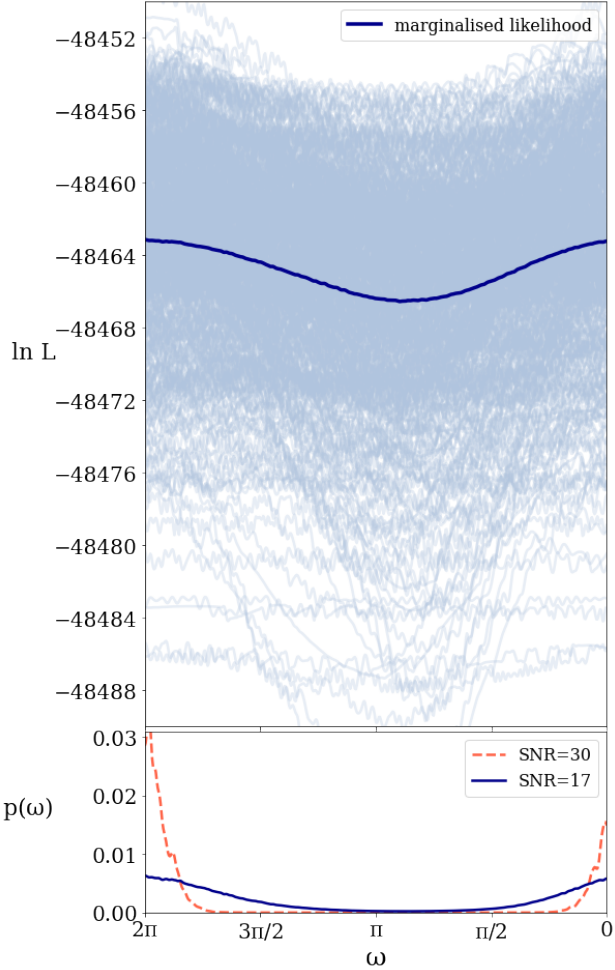}
    \caption{The argument of periapsis measured with \texttt{TEOBResumS}. Top: The likelihood over $\omega_\text{eff}$, plotted for 1000 posterior samples generated with \texttt{TEOBResumS}, with the averaged log likelihood highlighted in dark blue. Some of the curves appear almost completely straight, corresponding to posterior samples with eccentricity close to zero. Bottom: In dark blue is the posterior distribution over the same parameter space. In dashed red is shown the posterior obtained with the same injection but at SNR ($\approx$ 30). The width of the peaks of the distributions provide an indication of how well  we can resolve $\omega_\text{eff}$. $\omega_\text{eff}$ is beginning to become resolvable at SNR $\approx$ 17 but is highly resolvable at SNR $\approx 30$.}
    \label{fig:likelihood}
\end{figure}

Figure \ref{fig:likelihood} shows the posterior distribution for $\omega_\text{eff}$ ---our parameterisation for $\omega_\text{ref}$, calculated with \texttt{TEOBResumS}. We plot the results for the fiducial waveform (SNR $\approx$ 17) shown in Table~\ref{tab:teob_params} compared to the posterior obtained from a louder, SNR $\approx 30$ signal. 
The marginalised $\text{ln}(\mathcal{L})$ changes by $\approx$ 3.3 over the waveform cycle for the low-SNR injection. 
One rule of thumb states that a feature is strongly resolved if it is measured with  $\text{ln}(\mathcal{L})$ $\approx$ 8 \citep[e.g.,][]{jeffreys1998theory}.
With that threshold, our results indicate that we do not confidently resolve the argument of periapsis for this injection. At SNR $\approx$ 30, the $\text{ln}(\mathcal{L})$ changes by $\approx$ 20. This suggests that $\omega_\text{eff}$ is strongly resolved for this simulation.  
Figure \ref{fig:likelihood} (bottom panel) shows the posterior distribution for $\omega_\text{eff}$. The shape of the distribution suggests that $\omega_\text{eff} \approx 0 \pm \pi/2$ is moderately favoured by the data at SNR $\approx$ 17 and strongly favoured at SNR $\approx$ 30. The width of the peak is comparable to the prior volume, although values of $\omega_\text{eff} \approx \pi$ are quite strongly disfavoured. This means that while the data has found some preference for the value of $\omega_\text{eff}$, at  SNR $\approx 17$, it is not well constrained and is only moderately more informative than the prior probability distribution. At SNR $\approx 30$, the peak becomes narrower and rules out more of the prior volume - increasing the confidence of the measurement. As the SNR increases, the unfaithfulness of waveforms from numerical relativity simulations becomes more detectable along with $\omega_\text{ref}$. The mismatch from numerical relativity at $e \approx 0.1$ is $\approx 1 \%$ \citep{TEOBResumS, bonino_22}, which is less than the mismatch caused by $\omega_\text{ref}$ in Section \ref{sec:omega}. Hence, $\omega_\text{ref}$ is likely to be more important than waveform systematics at the SNR and eccentricities studied here. 

\begin{figure}
    \centering
    \includegraphics[trim = 15mm 15mm 15mm 15mm,width=\columnwidth]{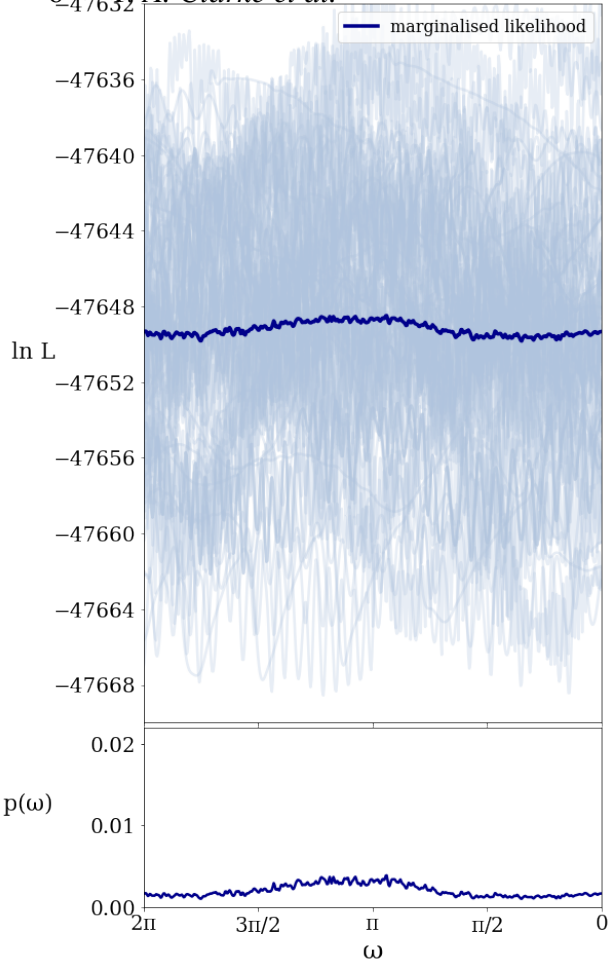}
    \caption{The argument of periapsis measured with \texttt{SEOBNRE}. Top: The likelihood over $\omega_\text{eff}$, calculated using the waveform \texttt{SEOBNRE}. The average is highlighted in dark blue. The  $\text{ln}(\mathcal{L})$ changes by $\approx$ 1.3, lower than for \texttt{TEOBResumS}. This result suggests that for this waveform, $\omega_\text{eff}$ can not be resolved in the data of eccentric binary black hole mergers at the SNR shown here ($\approx$ 17). Bottom: The posterior distribution for $\omega_\text{eff}$ over the parameter space. The posterior is noisier and less significant than the \texttt{TEOBResumS} result but has a similar width.}
    \label{fig:SEOB_likelihood}
\end{figure}

We repeat the SNR $\approx 17$ analysis with \texttt{SEOBNRE} and present an analogous version of Fig. \ref{fig:likelihood}, shown in Fig.~\ref{fig:SEOB_likelihood}. The results are similar to and consistent with those obtained using \texttt{TEOBResumS} ($\Delta \text{ln}(\mathcal{L}) \approx 1.3$), and support the evidence that $\omega_\text{eff}$ is not strongly resolvable in current eccentric gravitational-wave events. $\omega_\text{eff}$ is less resolvable in this simulation than in the analogous simulation with \texttt{TEOBResumS}. \cite{knee22} found that eccentricity values input to \texttt{TEOBRESumS} result in empirical eccentricities that are typically higher than for \texttt{SEOBNRE}. Our results seem consistent with this finding, since $\omega_\text{ref}$ seems to be more resolvable in \texttt{TEOBReumS} than \texttt{SEOBNRE}. This could be due to the differences in the waveform definitions of eccentricity. Future studies should compare waveforms with a ``Rosetta stone'' as in \cite{knee22} to account for different definitions of eccentricity. The traces along the $e_{10}$ path are much noisier than for \texttt{TEOBResumS}, producing a marginalised likelihood and posterior for $\omega_\text{eff}$ that is less smooth. Another difference between the results is that the \texttt{SEOBNRE} data prefer $\omega_\text{eff} \approx \pi \pm \pi/2$ which suggests the arbitrary reference argument of periapsis set by the fiducial waveform ($e_\text{fiducial} = 0.1, f_\text{fiducial} = \unit[10]{Hz}$) is out of phase by $\pi$ between the waveforms. The discrepancy is thought to be due to differences in the waveform systematics between the waveform models. In particular, in \texttt{SEOBNRE}, the reference frequency is subject to a corrective transform according to the reference eccentricity before constructing the waveform, which means that the waveforms generated might not follow the intended $e_{10}$ path. For more details on eccentric waveform systematics, see \cite{knee22}, Varma \& Pfeiffer., in prep.

There is a universal probability density function for the distribution of SNR for a population of binary black holes \citep{Schutz_2011, Chen_2014}, given by
\begin{equation}
    f_\rho = \frac{3 \rho{_\text{min}}^3}{\rho^4},
    \label{eq:snr_dist}
\end{equation}
where $\rho{_\text{min}}$ is the threshold SNR for a detection, assumed to be 12 for a 3-detector network. Hence, only $\sim$ 6 percent of events will be louder than SNR $\gtrsim$ 30. Since only $\sim$ 5 percent of mergers are expected to be eccentric at 10 Hz if dynamical assembly is the dominant formation channel \citep[e.g.,][]{ Samsing2017b, Samsing17, SamsingDOrazio18, Zevin:2020:channels}, this means that $\omega_\text{ref}$ will not be measurable in the vast majority of binary black holes with the current detector configuration. However, improved detector sensitivity and detection rates should mean that even 6 percent of eccentric events could become a substantial population, for which $\omega_\text{ref}$ should not be neglected. 

The complexity of stellar evolution and star cluster physics have made it difficult to predict the dominant binary black hole formation channels and distinguish them in gravitational-wave data. The orbital eccentricity of these systems is an important marker of the formation channels. However, to accurately infer the parameters of eccentric binaries, we need to consider the argument of periapsis, which is not inferred through parameter estimation currently.

In this work, we find that $\omega_\text{ref}$ becomes marginally resolvable with \texttt{TEOBResumS} while only beginning to be resolvable with \texttt{SEOBNRE} for a moderately eccentric binary black hole system (with parameters similar to GW150914) when the SNR exceeds approximately 17. By SNR $\approx 30$, $\omega$ becomes very well resolvable for the same system parameters. 
Given the modest SNR of current eccentric candidates (GW190521 was detected with SNR $\approx$ 15), past analyses that fix $\omega_\text{ref}$ to an arbitrary value are unlikely to suffer from significant systematic error.
However, as the gravitational-wave transient catalog grows, and more events are detected with higher SNR, it will soon become important to include the argument of periapsis in parameter estimation analyses. Future studies should consider marginalising over $\omega_\text{ref}$ to avoid introducing bias to the results.

At least four events in the current gravitational-wave transient catalogue may contain traces of eccentricity, including GW190521, GW190620\_030421 \citep{Romero-Shaw:2020:GW190521, Romero-Shaw2021, Gayathri_2022}, GW191109\_010717, GW200208\_222617 \citep{romero_shaw_22}, GW151226 and GW170608 \citep{Wu_2020, OSheaKumar2021}, GW190929 \citep{Iglesias_2022}. We show that, at least in the low-to-moderate eccentricity regime, the reference $\omega$ of these systems is not well resolvable. Higher-eccentricity injection studies are needed to determine the influence of the reference $\omega$ on the recovery of source parameters for systems with more extreme eccentricities. Therefore, it is likely that previous analyses that have not marginalized over $\omega_\text{ref}$ have results that are robust to changes in $\omega_\text{ref}$. Our recipe indirectly varies $\omega_\text{ref}$ by simultaneously adjusting $e_\text{ref}$ and $f_\text{ref}$. 
In the long-term, the only solution is to build waveform approximants that allow users to vary $\omega_\text{ref}$ directly.


\section{Acknowledgements}
We thank the referee for their helpful suggestions. This work is supported through Australian Research Council (ARC)  Centre of Excellence CE170100004, and Discovery Project DP220101610.  T. A. C. receives support from the Australian Government Research Training Program. I.M.R.-S. acknowledges support received from the Herchel Smith Postdoctoral Fellowship Fund. Computing was performed using the LIGO Laboratory computing cluster at California Institute of Technology, and the OzSTAR Australian national facility at Swinburne University of Technology.

\section{Data availability}
The data underlying this article will be shared on reasonable request to the corresponding author.


\bibliographystyle{mnras}
\bibliography{bib}
\end{document}